# Microsimulation of the Port of Dover

Chris Roadknight, Uwe Aickelin and Galina Sherman
Intelligent Modelling & Analysis Research Group (IMA)
School of Computer Science
The University of Nottingham
Jubilee Campus
Wollaton Road
Nottingham NG8 1BB
{cmr,uxa}@cs.nott.ac.uk
G.Sherman@2008.hull.ac.uk

***Abstract*— Modelling and simulating the traffic of heavily used but secure environments such as seaports and airports is of increasing importance. Errors made when simulating these environments can have long standing economic, social and environmental implications. This paper discusses issues and problems that may arise when designing a simulation strategy. Data for the Port is presented, methods for lightweight vehicle assessment that can be used to calibrate and validate simulations are also discussed along with a diagnosis of overcalibration issues. We show that decisions about where the intelligence lies in a system has important repercussions for the reliability of system statistics. Finally, conclusions are drawn about how microsimulations can be moved forward as a robust planning tool for the 21st century.**

***Keywords-Simulation, Transport, Operations Research, Agents***

## I. INTRODUCTION

Traffic microsimulation software is becoming increasingly complex, parameterized and configurable. Regardless of how graphically realistic the end product may appear, the core statistics generated by the simulation still needs to be validated and verified. Events and statistics that show up when simulations are tested must also appear in real life and vice versa. Real world validation of simulation results can be an expensive, time consuming, subjective and erroneous process and deciding exactly how much validation to commission is usually an imprecise art. Existing methods for micro validation such as number plate recognition and manual sampling are expensive and error prone. Weighing up the cost/reward ratio of validation is an important but non-trivial process.

Traffic Microsimulations use a discrete event [17] approach to the movement of vehicles over time where the behavior of a system is represented as a chronological sequence of events. Each event occurs at a unique instant in time, with each new instance of the system viewed as new state. They combine this with some degree of agent based behavior where elements within the simulation have a set of parameters and policies that they use to come to decisions. Agent based approaches are successfully used in traffic and transport management [7]. However, despite their suitability research in this area is not mature enough and moreover "some problem areas seem under-studied, e.g., the applicability of agent technology to strategic decision-making within transportation logistics" [7].

Microscopic traffic simulation models have unique characteristics because of their representation of interaction between drivers, vehicles, and roads. The increasing availability of powerful desktop computers has allowed sophisticated computer software to be used to model the behaviour of individual vehicles and their drivers in real time. Microsimulation can be applied to any scenario involving complex vehicle interactions and has been used to model roads, rail, air and sea ports [1, 2]. If validation is not properly performed, a traffic simulation model may not provide accurate results and shouldn't be used to make important decisions with financial, environmental and social impacts.

Microsimulation breaks a simulation down into the smallest sensible connected components. In the case of the simulation of a traffic scenario that would be vehicles and the smallest sensible stations (i.e. Toll booths, roundabouts, junctions, stop signs). Each micro-component needs to be accurately modeled but it is also important to correctly define dependencies and flows. It has been shown that questionable simulation predictions can result from a lack of dependencies that result from independently microsimulating elements of a larger simulation [14], this brings into question how best to validate a simulation made up of large numbers of subcomponents and how to ensure the simulation does not contain small but significant errors.

Many scientific search and optimization approaches have analysed the subject of overtraining and the resulting lack of generalization. For instance, neural networks usually have a stopping condition which when reached signals the end of training, beyond this point the representation model continues to improve on the training subset of instances but  decays when tested on an unseen dataset [15]. A similar situation occurs in statistics when a statistical model starts to describe random error or noise instead of the underlying relationship, here it is called overfitting [8]. Overfitting or overtraining is more likely to occur when a model is unnecessarily complex, such as having too many parameters relative to the number of observations. While less well researched, similar situations may arise in simulations where a simulation is constructed to such an accuracy as to completely mimic the situation used as an example. It is easier to create overcalibration errors using modern, componantised simulation software where each individual element can be highly configured to be representative of the isolated sub-system without requiring any system wide validity.

The research in this paper involves simulations and real worth data from the Port of Dover. It was chosen for this research as it is the most important trading route between the UK and mainland Europe, has an intricate and multilevel layout (Figure 1) and has a substantial amount of existing data and simulations. Over the past 20 years, the number of road haulage vehicles (RHVs) using the Port of Dover has more than doubled to over 2.3 million [4]. Looking ahead over the next 30 years, both the Port and UK Government have forecast substantial growth in RHV freight traffic. Approximately 3 million tourist vehicles also pass through the ferry port annually making it a key European and global tourist gateway.

This paper sets out to identify the performance and characteristics of a microsimulation approach to closed system vehicle simulation with particular reference to the stability and reproducibility of the simulations. The next section of this paper outlines existing data, statistics and graphs for the Port of Dover, section three discusses the simulation software package VISSIM, section four introduces a novel validation procedure which is tested at the Port of Dover, Section 5 discusses an extension of the validated simulation and Section 6 offers some conclusions.

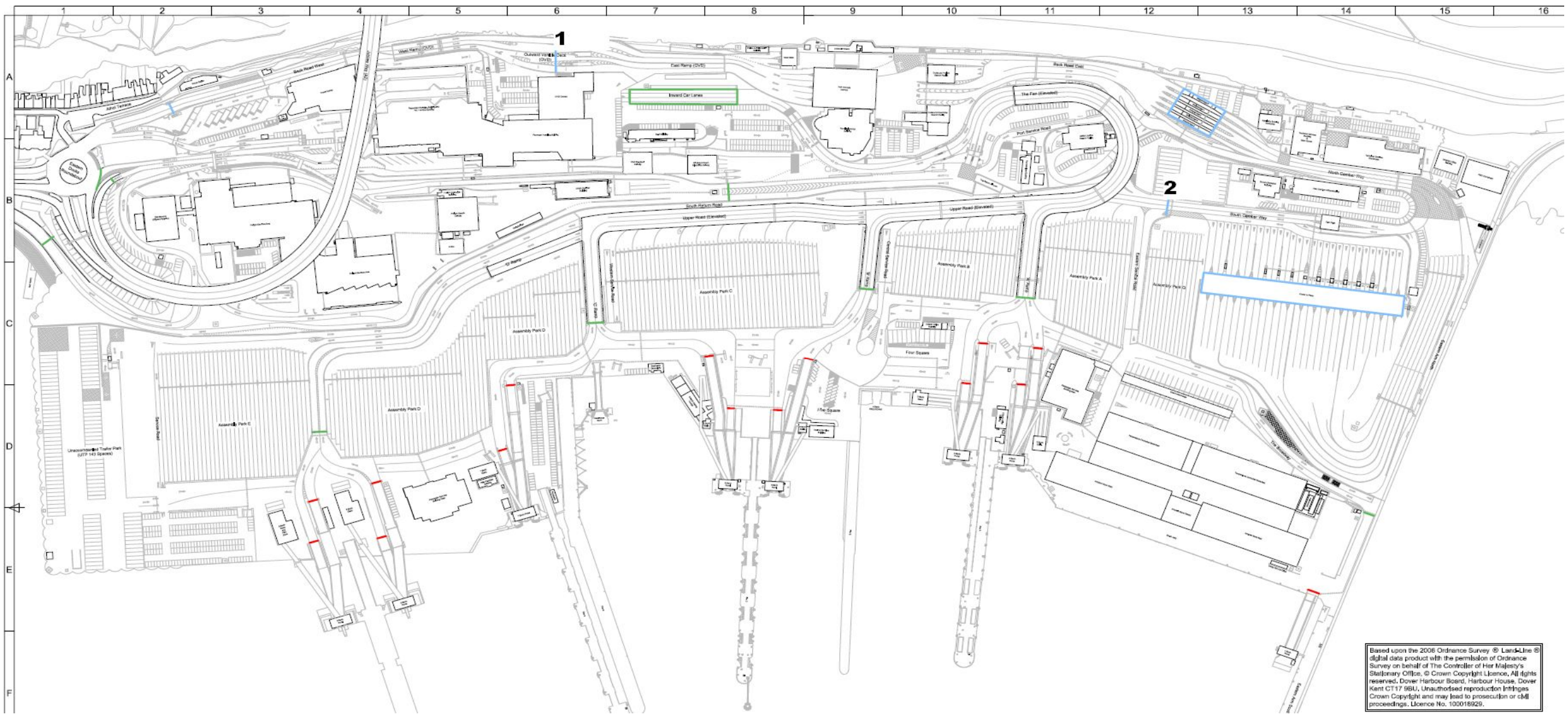

**Figure 1: Plan of the Port of Dover**

## II. Dover Existing Data

Looking at the RHV traffic for the 24 hour cycle over a full year it is apparent that systematic flow variations occur, Figure 2 highlights some key facts about this flow for 2009. For instance the maximum RHV flow is approximately four times the minimum flow at between one and four vehicles per minute. The lowest flow was between 2:00 to 2:15am and the highest flow between 3:15 to 3:30pm. These measurements were taken at the weighbridge, here all RHVs are weighed, timestamped and the driver side of the vehicle noted. This showed that 1,194,973 RHVs exited the UK via the port in 2009, of these 960,878 were left hand drive. The arrival statistics at the weighstations is related to the arrival at the port in general but modified at peak periods due to the first bottleneck at the port, the Customs check, where passports are inspected and also by the queueing at the weighbridges themselves. This initial check has the effect of smoothing the flow to the weighbridge and ticketing kiosks because at peak times queues build up, this effectively reduces the bursty-ness of the traffic flow. This doesn't change the overall numbers going through the port, just the arrival dynamics.

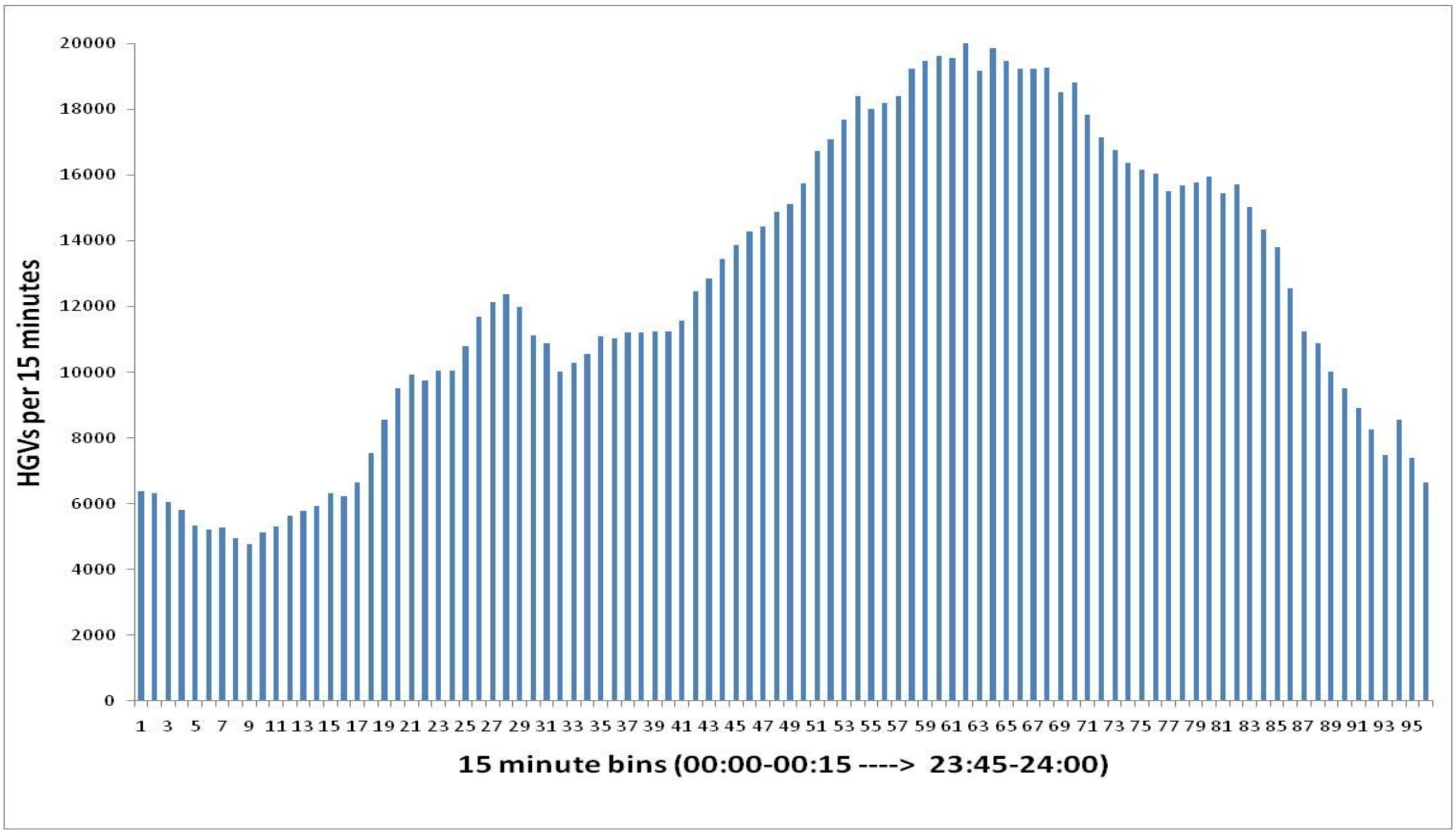

**Figure 2: Aggregated RHV flows for 15 minute time windows for the whole of 2009**

One way to measure actual arrival rates at the port is to use CCTV camera images to capture individual arrival of vehicles (Figure 3). The flow at these points can be automated or manually assessed. An example of the arrival process at the first bottleneck (passport check) is shown in Figure 4. When these values are plotted as a cumulative curve (Figure 5) the slope of the curve is a good indicator of arrival rates. We can also see the arrivals at one of the tourist check-in kiosks, Figure 6 shows these in 10 second bins.

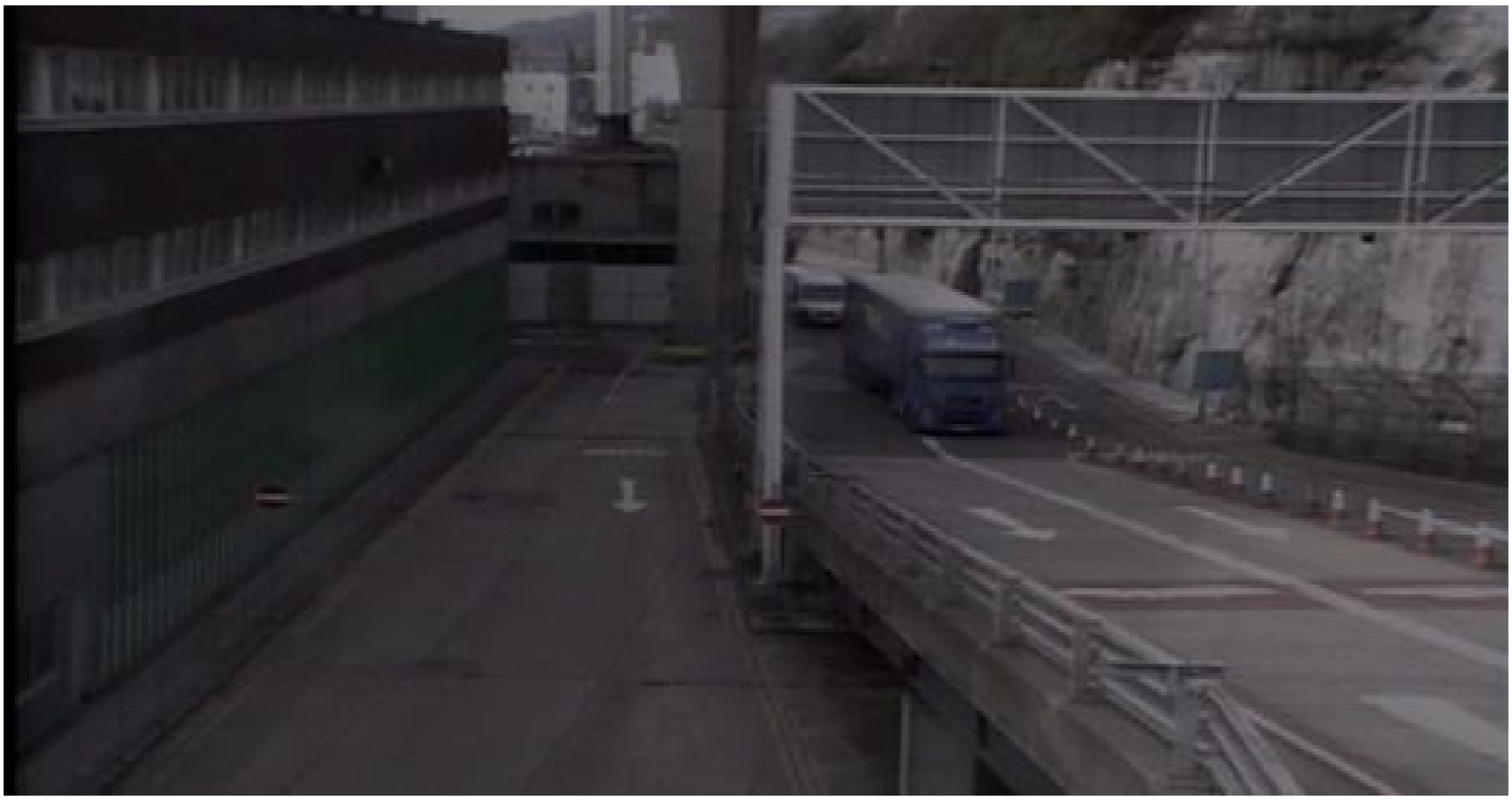

Figure 3: CCTV of vehicles arriving at the port

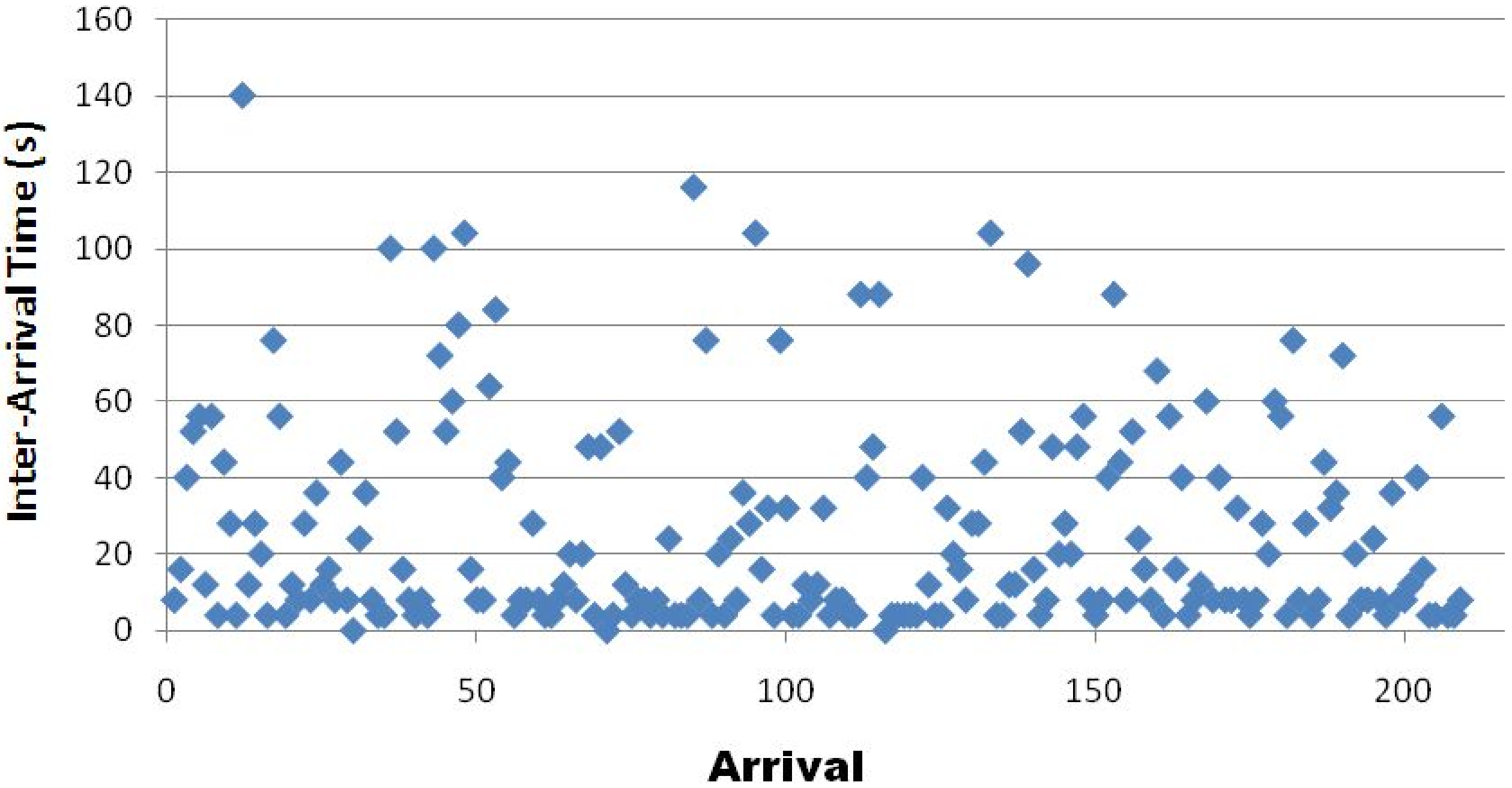

Figure 4: Sample of interarrival times of vehicles at the Port of Dover

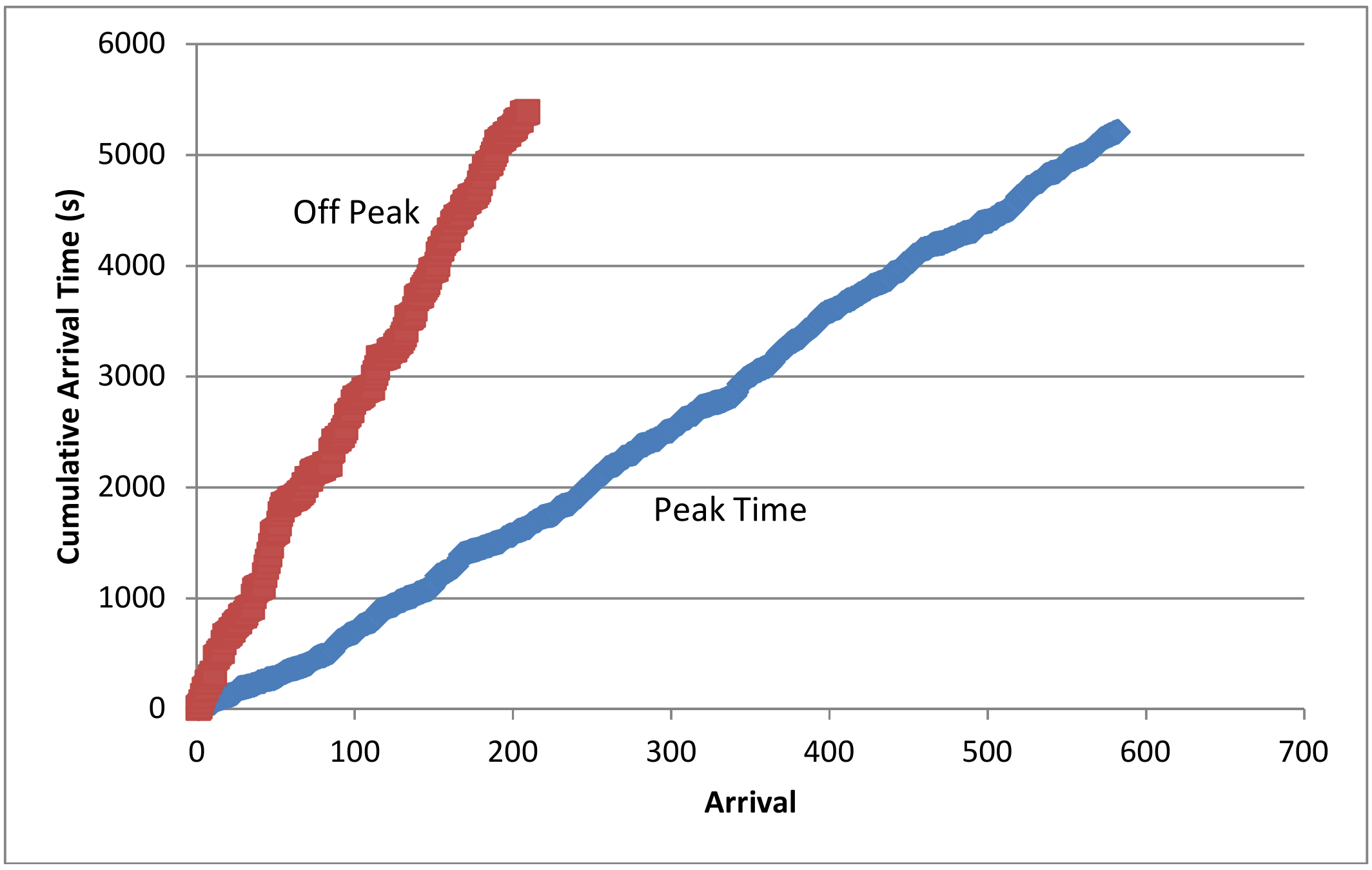

**Figure 5: Cumulative arrival times at peak and off-peak times**

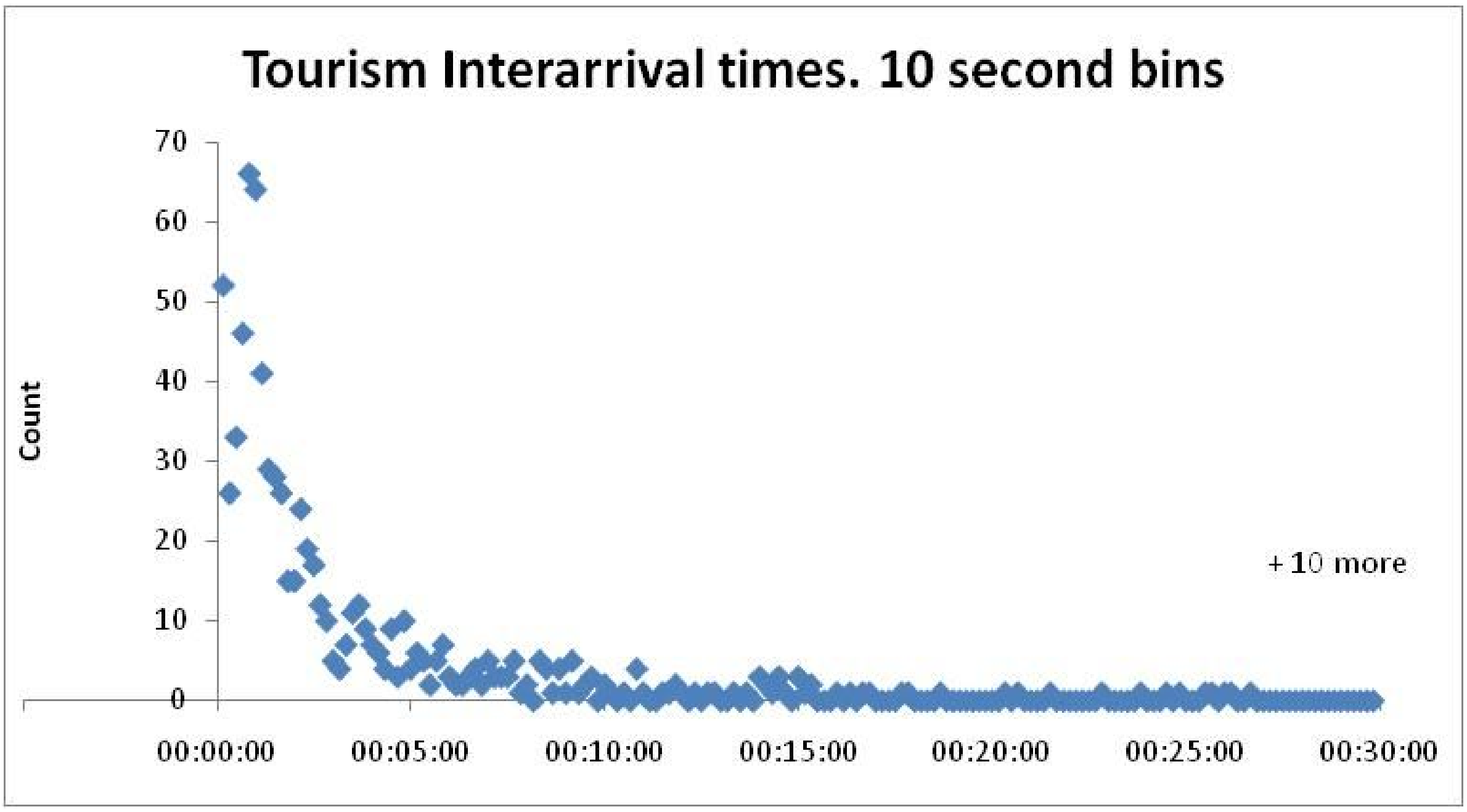

**Figure 6: Interarrival times at tourist check-in in 10 second bins for two days. Note: 10 interarrival times of greater than 30 minutes not displayed on graph**

## III. SIMULATION

Traffic simulations of transport networks traditionally use a discrete event, cellular automata style approach. Examples of this include TRANSIMS [9], PARAMICS [10], CORSIM [11] and more recently VISSIM [12]. Figure 7 shows an example of what a VISSIM simulation of the Port may look like. VISSIM [3] is a leading microscopic simulation program for multi-modal traffic flow modelling. It has a high level of vehicle behaviour detail that can be used to simulate urban and highway traffic, including pedestrians, cyclists and motorised vehicles. It is a highly parameterised design system that allows a lot of flexibility.

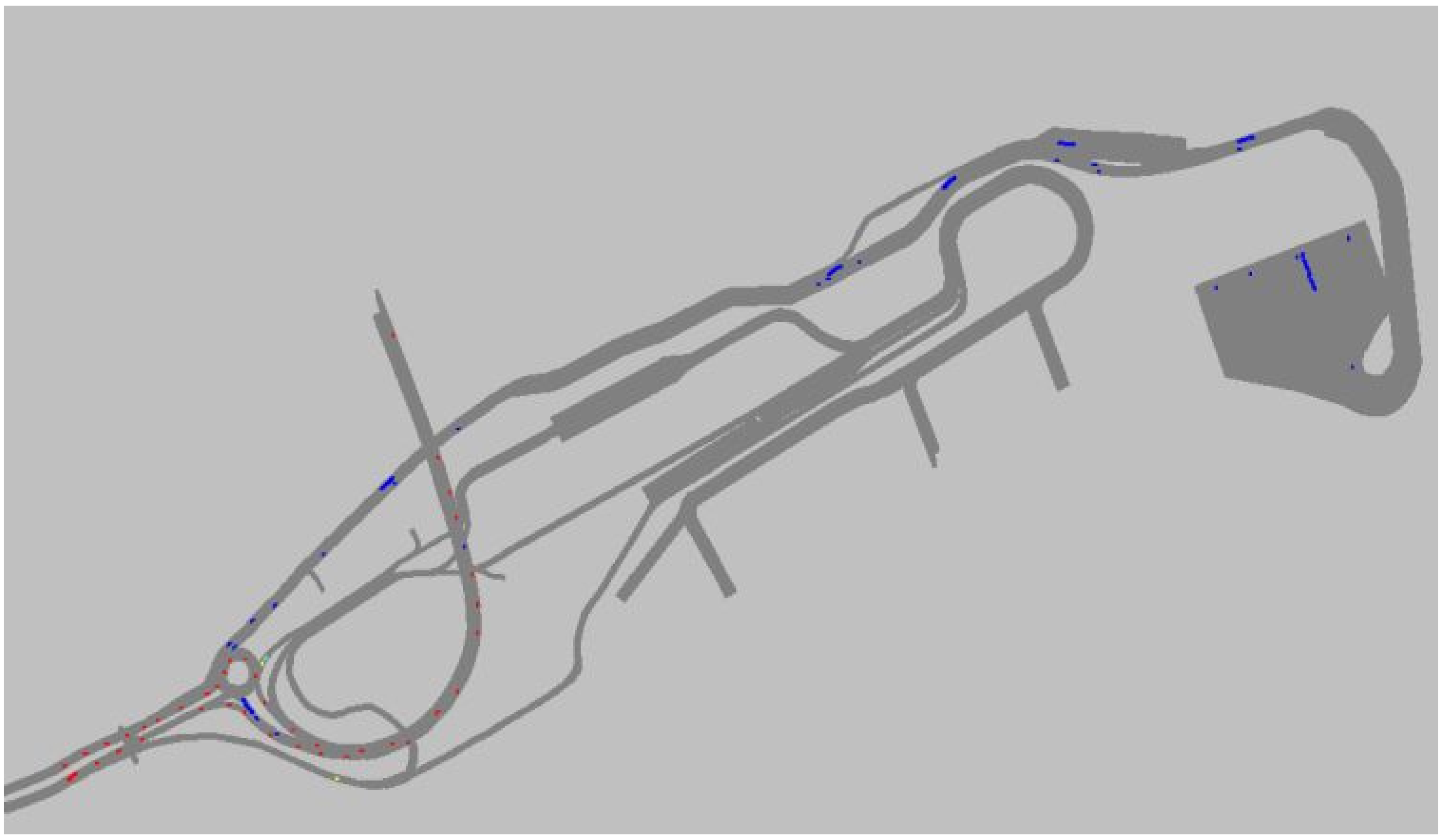

**Figure 7: VISSIM simulation of the Port of Dover**

VISSIM models provide detailed estimates of evolving network conditions by modelling time-varying demand patterns and individual drivers' detailed behavioural decisions [3]. Several model inputs (such as origin flows) and parameters (car-following and lane-changing coefficients) must be specified before these simulation tools can be applied, and their values must be determined so that the simulation output accurately replicates the reality reflected in traffic measurements.

There are several significant choke points around the port, places where queues appear and significant delays can arise, namely the passport checking area, the RHV weighbridge and the ticketing booths. Delays can also be introduced with additional security checks to a randomly selected percentage of vehicles. There are five weighbridges that all RHVs must stop at, RHVs are guided into the left two lanes coming off the Eastern Docks roundabout feeding into the three left most customs channels as to not impede the flow of other vehicles into the port. The wait time at the weighbridge is modeled as a normal distribution with a mean of twenty seconds and standard deviation of two seconds.

VISSIM allows the specification of an initial random number seed, this allows for the same simulation to be repeatedly stressed with a different sequence of random numbers but also allows direct comparisons of different scenarios using the same random numbers. Variability between runs with different seeds is a good metric for how robust the system is. Large differences in run statistics when using different random numbers suggests one of two things.

- Some kind of chaotic data/environment interaction.
- Some kind of illogical and pathological fault in the simulation design.

Each section of road, link, junction etc has to be accurately modelled. The simulation might develop an inbuilt fault whereby a small design aspect that appears (on some levels) to be sensible produces considerable variation in validation statistics just by modifying the random number seed. For instance, an integral component of traffic simulations is the decisions made by drivers as they navigate the desired road sections. Lane selection, overtaking, acceleration, deceleration, follow gaps are all examples of driver behaviour parameters. The port of Dover has many lane selection points, and the number possible of lanes changes repeatedly. By monitoring the lane usage we can see the desired occupancy rates of lanes but configuring the system to correctly reflect this is non-trivial. For instance, we know the occupancy of the five weighbridges over the whole of 2009 was for bays one to five were 22.3%, 25.4%, 22.8%, 15.6% and 13.9% respectively.

One way to enforce this ratio is to use a probabilistic, ‘roulette wheel’ style lane selection policy. VISSIM, along with most simulation toolkits, offers methods to specify probabilistic routing whereby a defined percentage of vehicles are sent down unique routes. This is a piecewise technique that can be reapplied at various locations around a simulation. While these methods are attractive from a calibration perspective as exact representations of existing statistics can be ensured, the process is an unrealistic one as it assumes that drivers make probabilistic decisions at precise locations. So in this case when a vehicle arrives at a point prior to the weighbridges it is allocated one of the lanes based on the respective probabilities. It turns out that this method leads to significant variations in trip times depending on the initial random number seed (Figure 8). One of the benefits of graphical microsimulation is that the 2D and 3D simulations help the researcher to visualise a new scheme and its potential benefits but also to highlight unrealistic behaviour. Figure 8 shows the congestion at the decision point for two different runs. Using probabilistic routing to enforce correct routing percentages is a clear case of overcalibration affecting simulation brittleness.

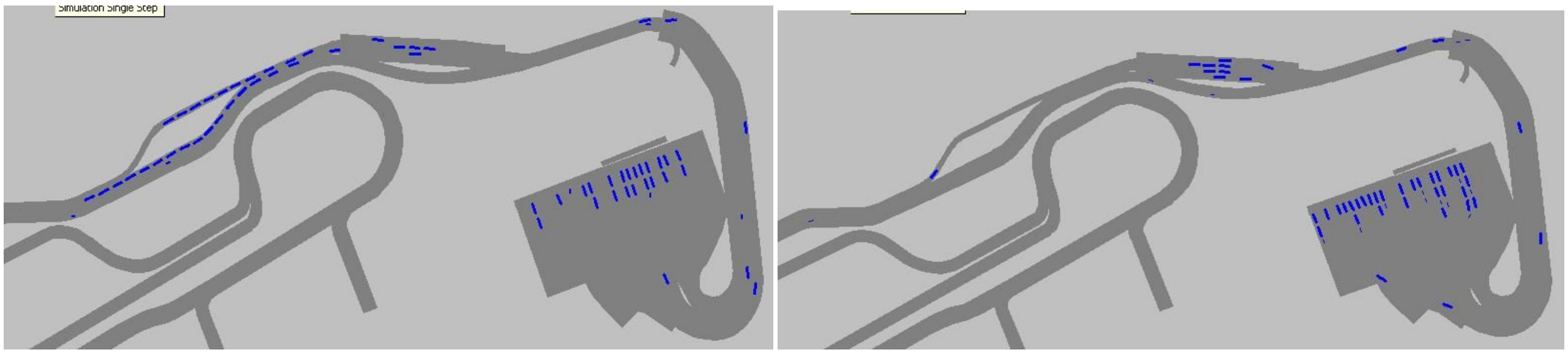

**Figure 8: VISSIM simulations at identical times, with identical traffic flows but different random number seeds showing how this can effect congestion if probabilistic routing is used**

These runs have identical and realistic inbound traffic flow rates that have been constructed based on observations of flows at peak rates, yet considerable difference is behaviour. The flows were generated by recording arrival at the port using CCTV camera footage and constructing an arrival process based on two minute segments. Each two minute segment produced exactly the number of vehicles required (Figure 9), vehicles enter the link during that two minute period according to a Poisson distribution, if the defined traffic volume exceeds the link capacity the vehicles are stacked outside the network until space is available again but this was not required in the Dover simulation. Figure 8 shows how one busy 90 minute period was represented. Figure 10a shows the queue lengths at the weighbridge for the two simulation runs where the only difference is the random number seed, differences in queue lengths of up to 140 meters are apparent. Figure 10b shows the associated trip time differences between these two runs with average trip times for all vehicles varying by 100% between the two simulations during some points of the simulation. The probabilistic approach has a lack of flexibility where drivers will stick to their random number allocated lane even if it is congested. We found much more repeatable results could be gained by allowing VISSIMs inbuilt driver behaviour features to select the best lane. Even though the desired percentage occupancies were not enforced, similar weighbridge ratios were generated as simulated drivers avoided the congested centre lanes at peak times and effect of random number seeds was much less.

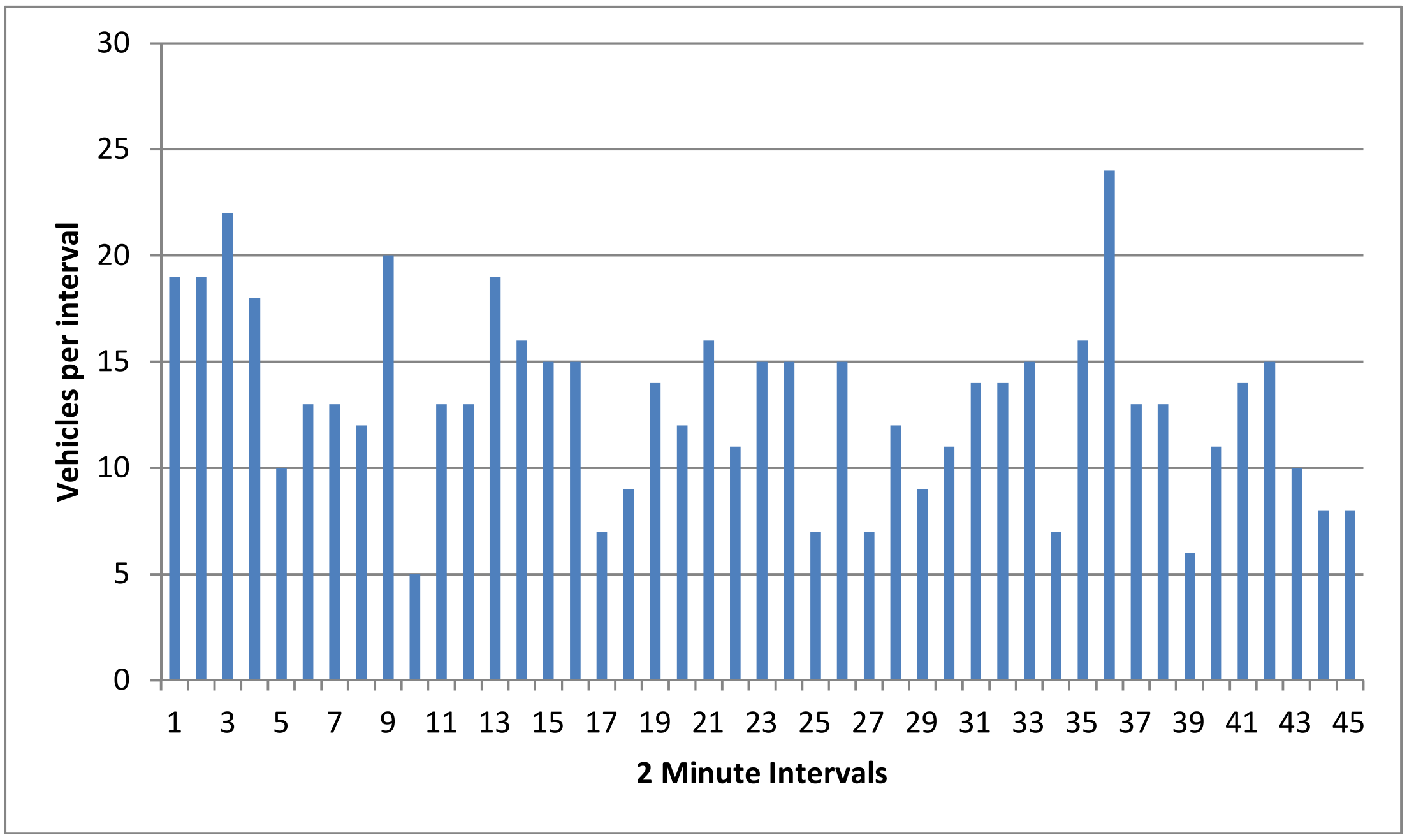

Figure 9: Typical arrival process in two minute intervals

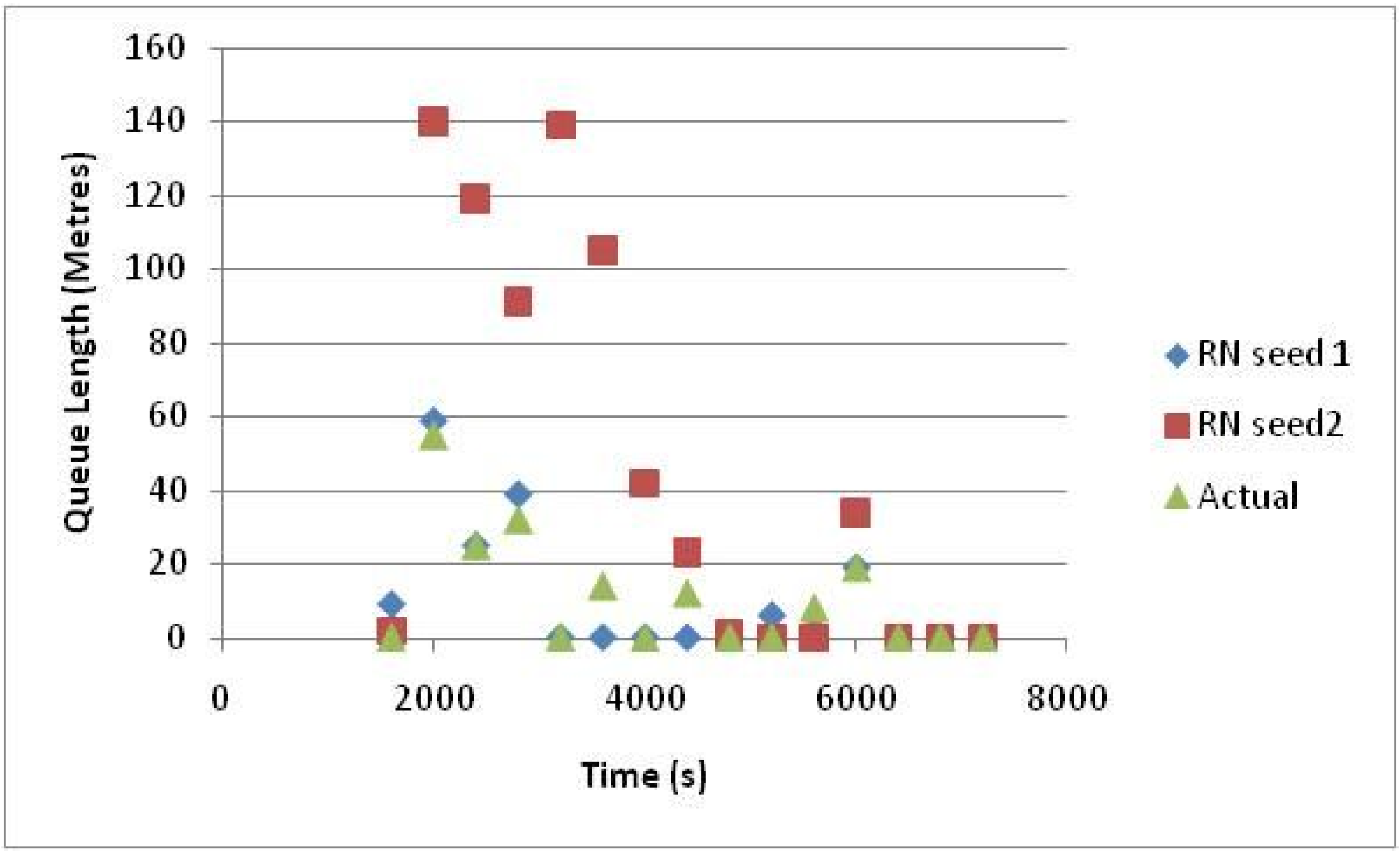

Figure 10a: Queue lengths for identical flows but different random number seeds

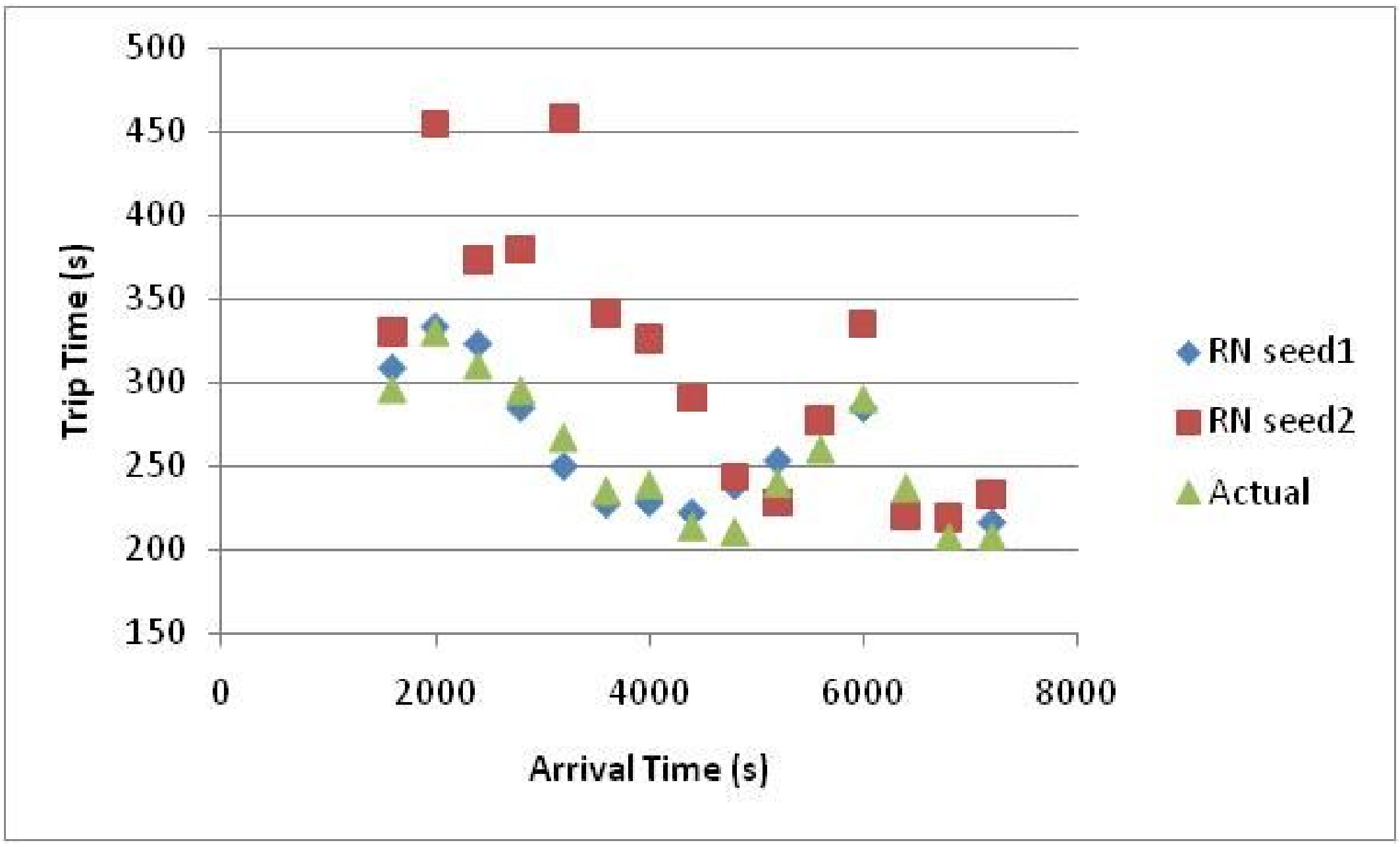

**Figure 10b: Effect of random seed on trip time**

## IV. Decision Making in Simulations

All simulations require decisions to be made, frequently and at many locations. For vehicle simulations these decisions relate to the behaviour of drivers and their vehicles in response to environmental stimulus and their own goals. An agent based solution would migrate these decisions to each individual driver and vehicle. This approach is sensible and realistic but extremely difficult to manage and configure, there are also computation issues when decisions are not aggregated. Decisions can be made on a higher level and most modern simulation software systems allow for methods such as conditional, probabilistic and deterministic routing. The benefit of aggregated high level decision making is that calibration and configuration is much easier to manage and enforce, for instance if we know that 30% of vehicles take one route and 70% take another route then a simple random or round robin selection procedure would ensure a near exact reflection of this in the simulation. Achieving precise behaviour statistics using a low level agent approach is much more difficult.

The Port of Dover is an ideal example of the difficult trade-offs for where the decision making processes should be located. Firstly there is the driver specific decisions that affect how their vehicle behaves, how much space do they leave to the vehicle in front, when do they overtake, how hard do they brake/accelerate etc. There are also lane choice decisions that are made, how quickly these decisions are made, where they are made, how aggressive etc. In this section we contrast two methods of decision making and how these methods differ in terms of accuracy, particularly under different loads.

The weighbridge is a key point for the flow of vehicles through the Port of Dover. Here every RHV must come to a halt to be weighed, and every driver chooses from five lanes. The lane selection data for the whole of 2009 was made available so we had excellent statistics for how which drivers chose which lanes at which times. So when making the decision on which lane to choose we assessed two options:

1. Probabilistic Routing. At a set point prior to the weighbridge a random number is generated and a biased roulette wheel selection approach [16] is used to tell the driver which lane to head for. So if we knew 10% of drivers used lane four we could generate a random number between zero and one and if the number was greater than zero and less than or equal to 0.1 then the driver would select lane four. The ratios of vehicles to lanes is different at different flow rates, so a coarse approach would be to use the average lane section ratios for the whole one year period for all flow rates (Non-flow specific probabilistic routing) and a more precise method would use flow specific lane occupancy ratios (Flow specific probabilistic routing)

2. Agent based Routing [7]. Each driver has a set of configuration parameters that decides when they should overtake, change lanes, slow down etc. These configurations, along with the road layout and volume of traffic would be allowed to dictate the flow of vehicles through the weighbridge

We use five different flow rates in these experiments, these are based on real world data extracted from video images of vehicles entering the Port and the timestamped weighbridge data. The rates range from 30 to 800 vehicle an hour. All graphs show average lane selection or trip time statistics captured over 21 replications and error bars based on standard deviation or quartiles where appropriate. Figure 11 shows an overhead plan of the weighbridges with lane numbering, with lane five being nearest farthest from the sea and lane four the nearest.

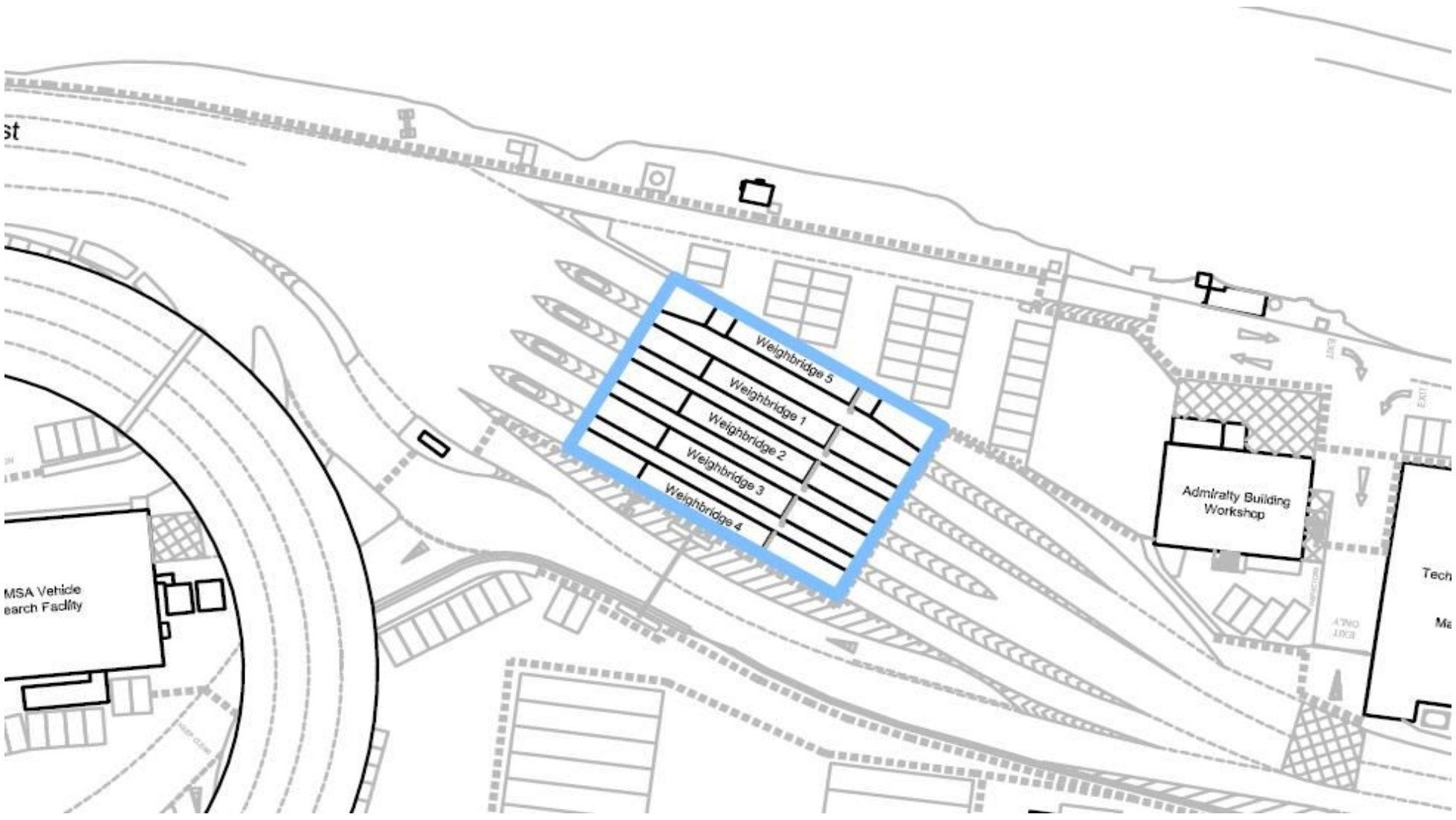

**Figure 11: Weighbridge Plan with Numbering**

Figure 12a shows the lane occupancy observed for the agent based simulation approach. At low flow levels lanes one and two make up nearly 100% of lane usage, this is a result of the low occupancy of lanes meaning the drivers seldom need to change lanes, this is shown in more detail in Figure 12b. As the flow increases all five lanes are used more, as queues for lanes one and two develop. Figure 13 show what was actually observed at the weighbridge, again use of lanes one and two predominate but to a less marked degree. Figure 14 shows when the agent based simulation error for each of the five lanes is aggregated and compared with the error for two flow based approaches. It shows that at very low flows the agent based approach has a significant error but at very high flow rates is more accurate than the probabilistic routing using average flows. While in this instance it could be surmised that the performance of the simulation at high flow is more important this will not always be the case for all simulations. Correcting the agent based approach would involve introducing more complex decision making for the drivers and capturing the reasons that they choose the lanes they choose. Correcting the flow based decision making process is just a case of using flow specific statistics when parameterising the probabilistic route selection

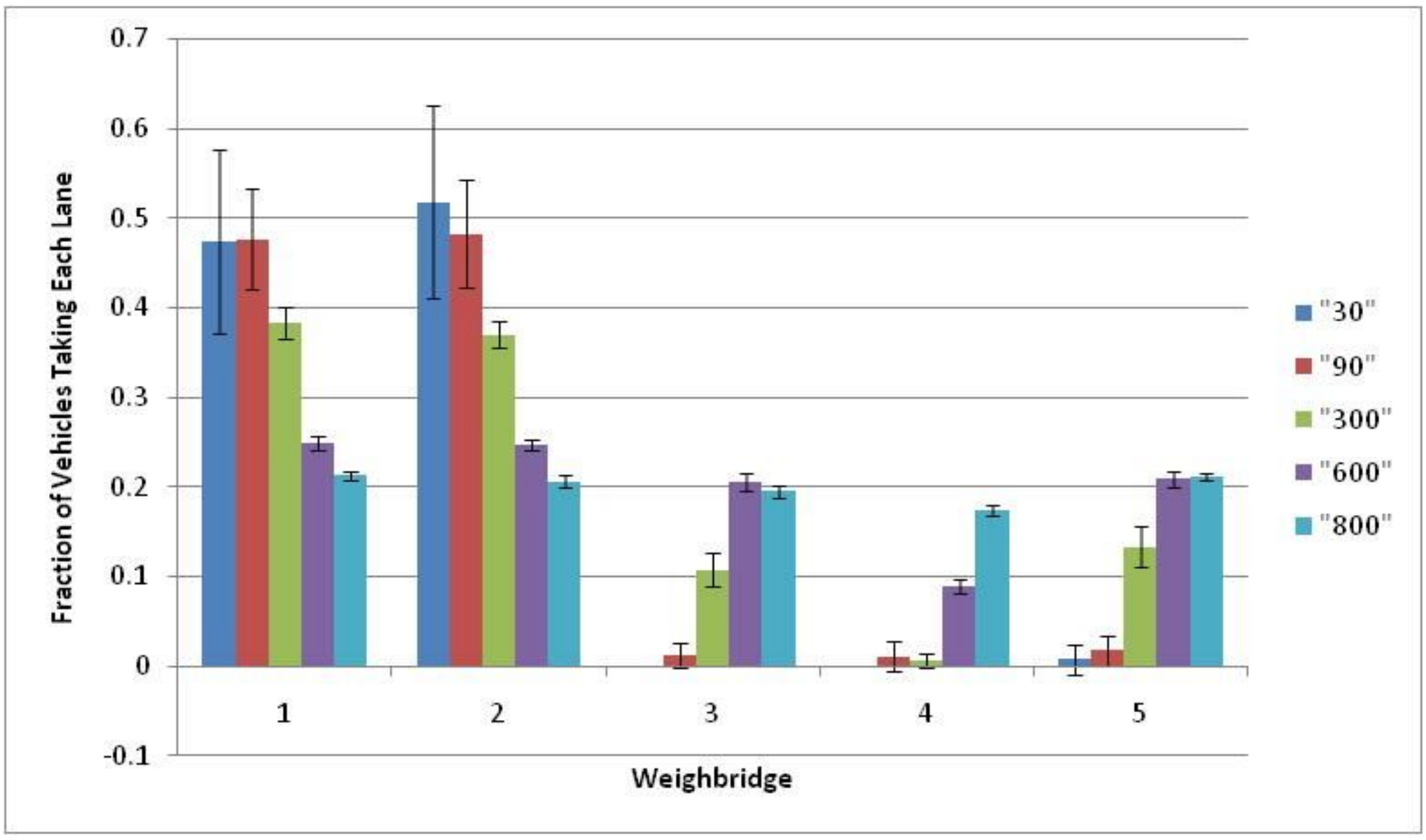

**Figure 12a:** Lane occupancy ratios at various arrival rates for the agent based simulation approach

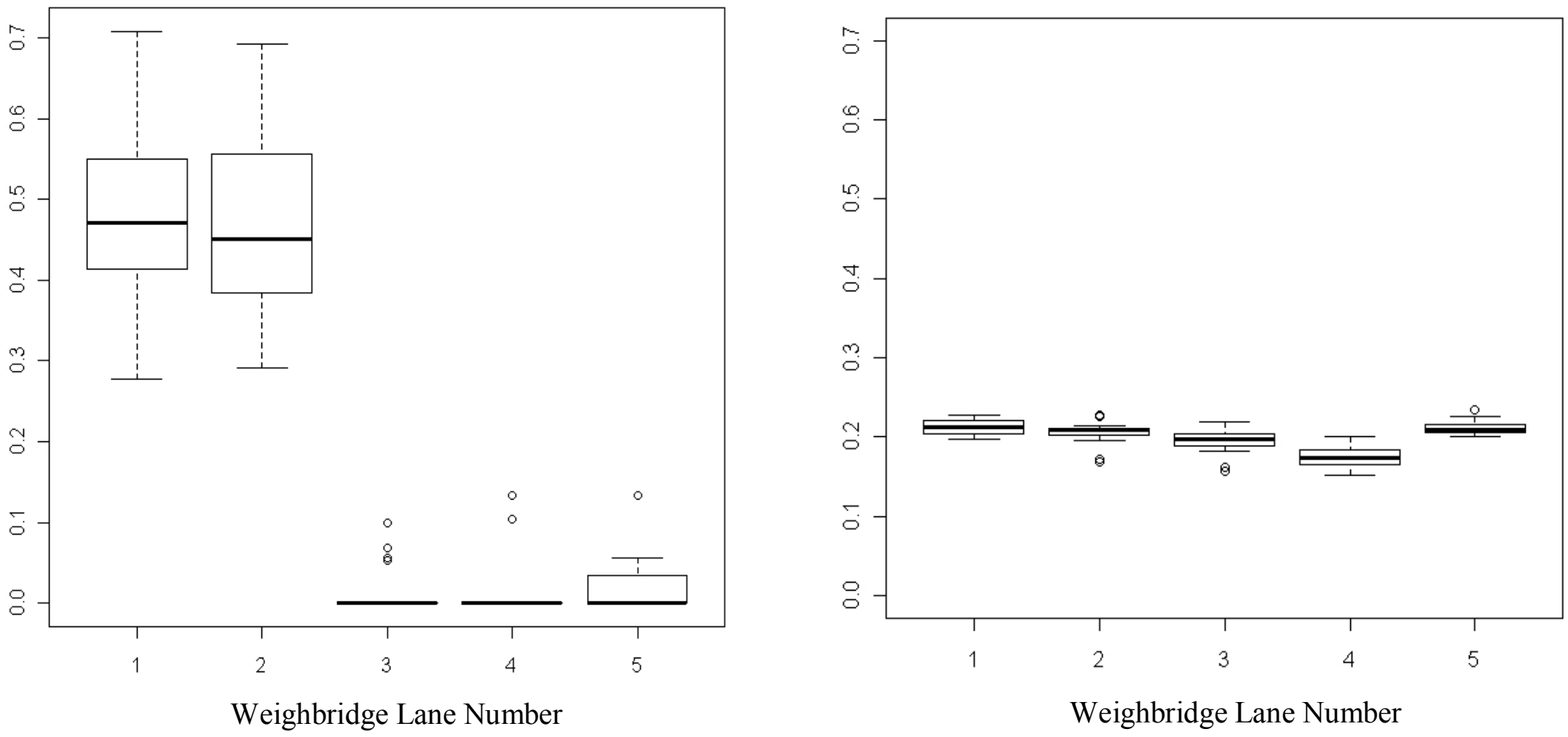

**Figure 12b:** Box and Whisker plots of the weighbridge lane selections made in the agent-based simulation for a low flow of 60 vehicle per hour (left) and a high flow of 800 vehicles per hour (right).

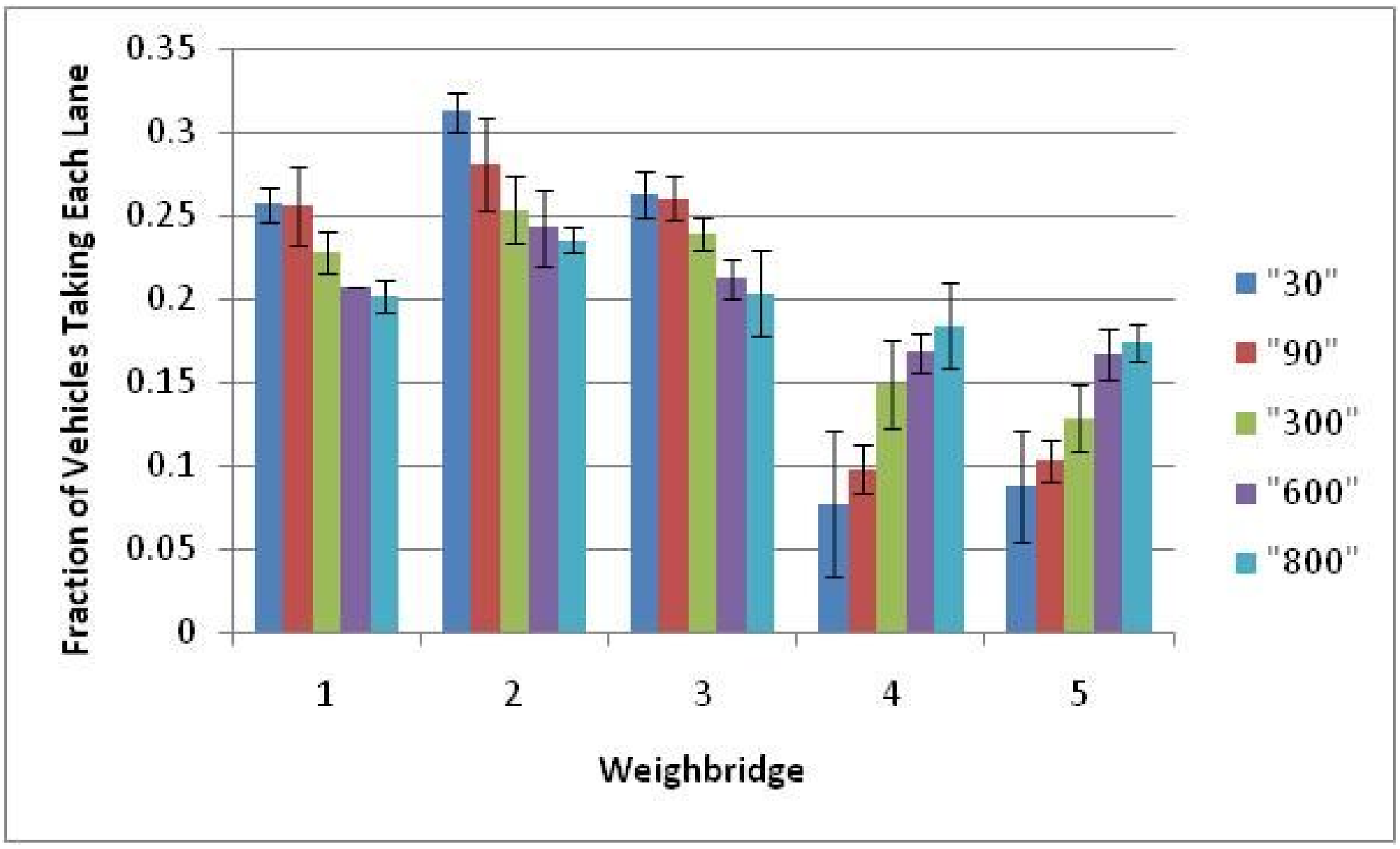

**Figure 13a:** Real World lane occupancy rates at various arrival rates

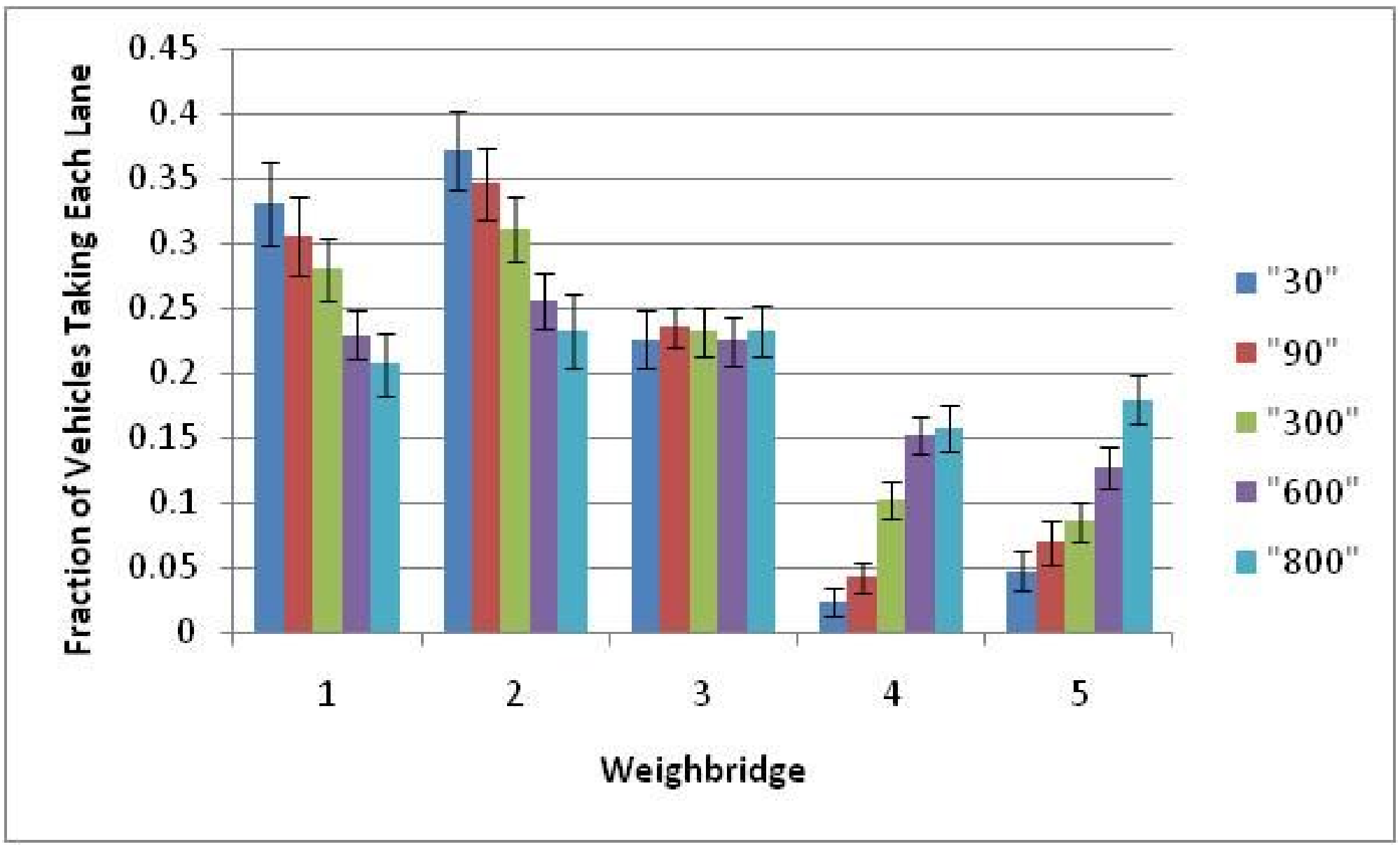

**Figure 13b:** Simulated lane occupancy rates at various arrival rates using probabilistic routing

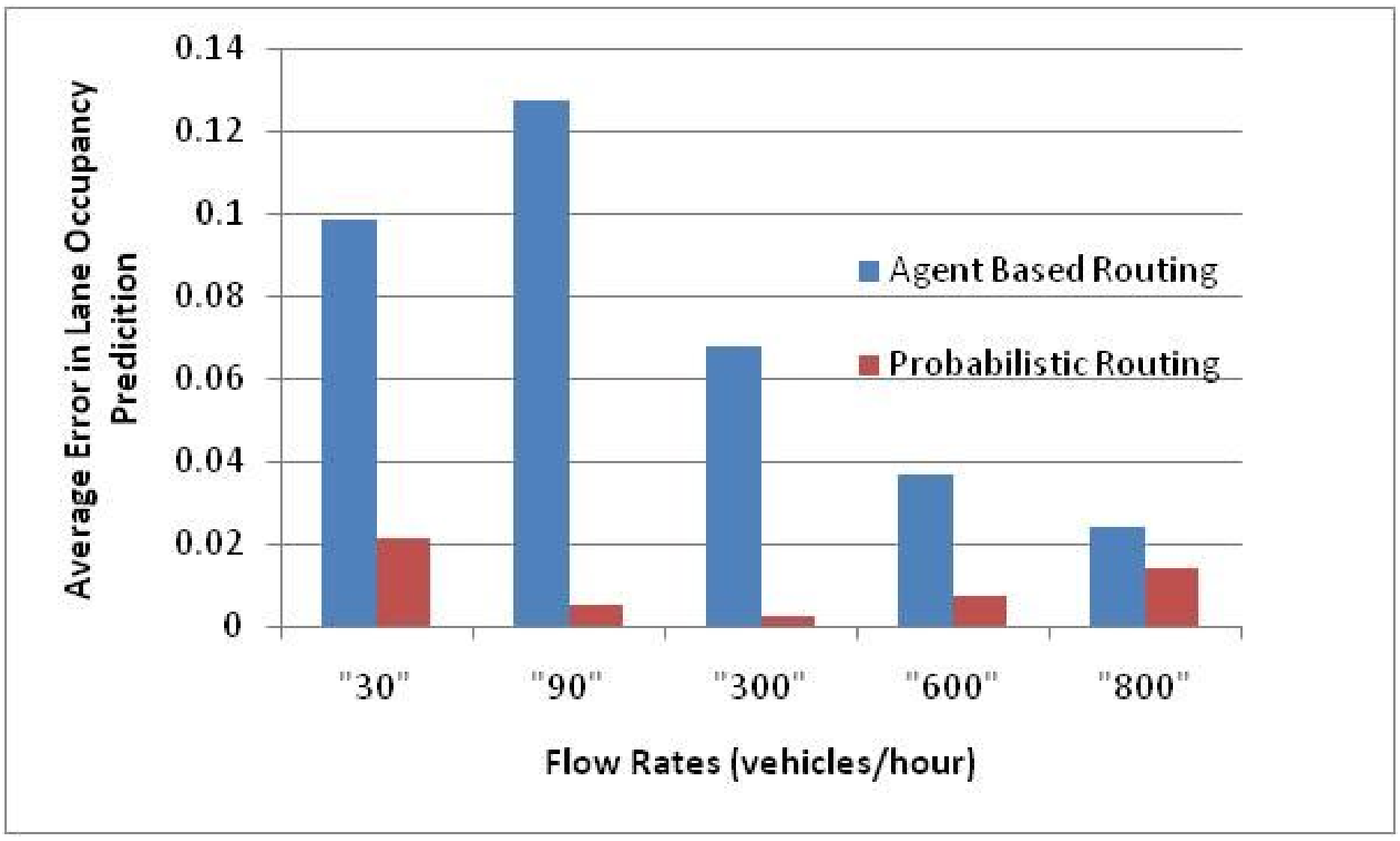

**Figure 14:** Errors in lane occupancy prediction for various simulation methods

While the correct reflection of real world occupancy is an important goal for this simulation, it is also important to check other metrics of evaluation, one such metric is the trip time or how long it takes for RHVs to travel between two points of the Port. While the last section shows how closely the weighbridge occupancy statistics can be engineered, this experiment will show how well the different simulation methods reflect the real point to point trip times, regardless of how well the weighbridge selection is performed. For this experiment we use a much shorter trip distance that starts after the customs check and continues until shortly after the weighbridge, this captures the delay caused by congestion at and approaching the weighbridge and discounts the effects of the ticketing and customs stop points. Figure 15 shows how both the agent based and probabilistic routing approaches are accurate reflectors of trip time at low and medium flow rates but also shows how the probabilistic routing approach gives much higher predicted trip times than those actually observed at high flow rates. This is due to an inflexibility of the probabilistic routing approach whereby once the desired lane has been selected the driver has no ability to over-rule this dictate and may cause considerable and unnecessary congestion by pursuing the require lane. Table 1 gives an overview of how much error is associated which each simulation method when measured again trip time and lane occupancy. The two largest errors (X) are the agent based approach performing poorly at low flow rate lane selection and the probabilistic routing approach performing poorly when trip times are measured at very busy times. The performance of both of these scenarios could be improved by introducing more driver intelligence but not without a significant increase in complexity and simulation run times.

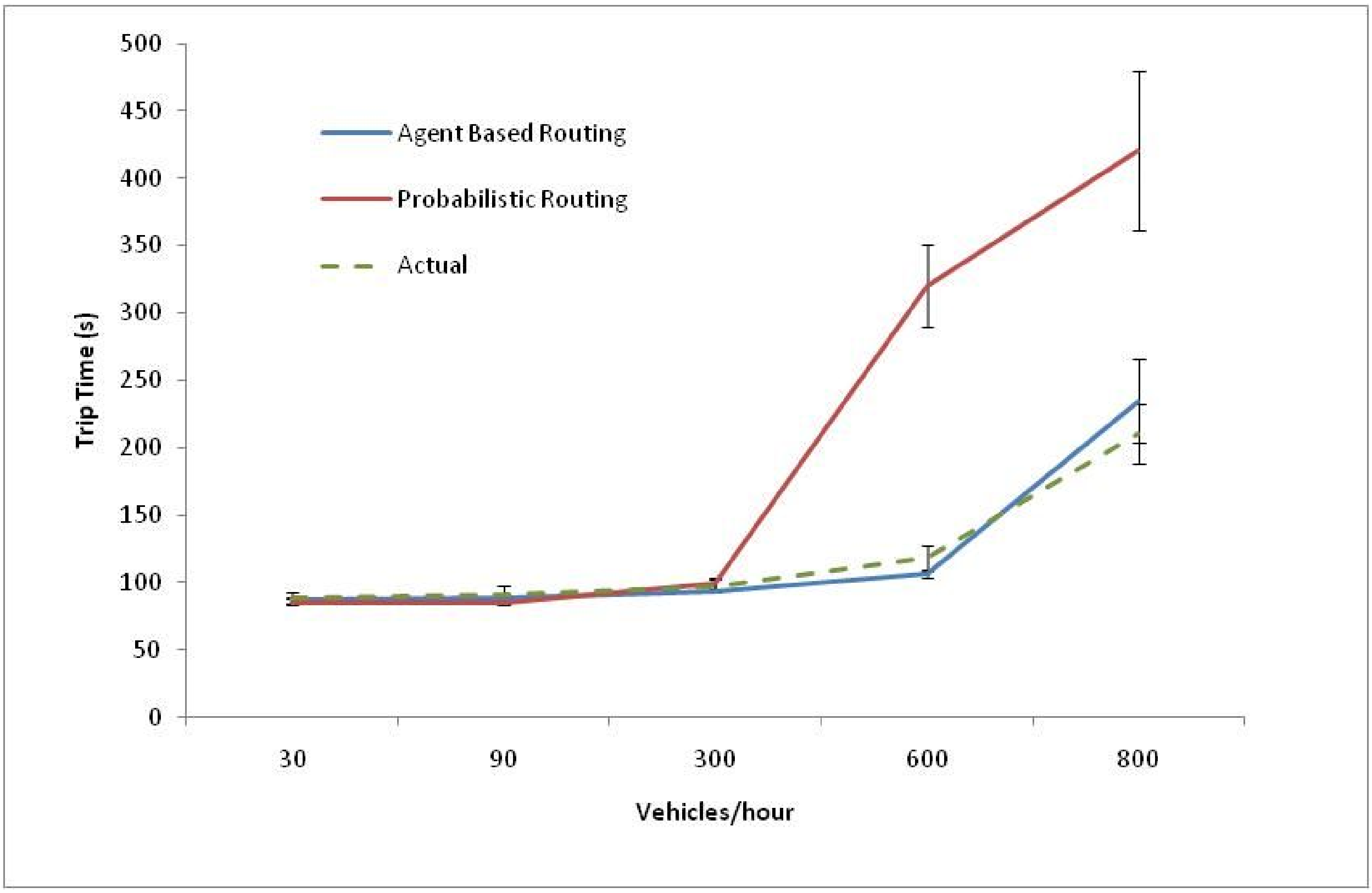

**Figure 15:** Trip times through the weighbridge complex at different arrival rates for real data and simulation methods

Arrival Rates

| | Low (30-90 vehicle/hour) | Medium (300 vehicles/hour) | High (600 vehicles/hour) | V. High (800 vehicles/hour) |
|---|---|---|---|---|
| Agent (Lane Selection) | X | ✓ | ✓✓ | ✓✓ |
| Probabilistic Routing (Lane Selection) | ✓✓✓ | ✓✓✓ | ✓✓✓ | ✓✓✓ |
| Agent (Trip Time) | ✓✓✓✓ | ✓✓✓✓ | ✓✓✓ | ✓✓✓ |
| Prob Routing (Trip Time) | ✓✓✓✓ | ✓✓✓✓ | ✓ | X |

**Table 1:** Performance of two simulation methods, using two metrics and four arrival rates

## V. VALIDATION

Validating microscopic traffic simulation models incorporates several challenges because of the incompleteness and rareness of validation data. Validation data is also usually measured in aggregate forms and not at the level of the individual vehicle. The cost-performance relationship of validation is an important function that should be well understood and used when deciding what extent any validation should be taken too [13]. Most of the model validation research uses average link measurements of traffic characteristics [6]. However, these approaches have limitations including possible non-obvious inconsistency between the observed and simulation estimated variables. Here we decided to use passive Bluetooth [5] monitoring to sample a vehicles location by uniquely identifying it using their Bluetooth signal id.

Two sampling locations were chosen near the beginning and end of the drivers' trip through the Port of Dover, from entry to the Port to waiting to embark on the ferry. These sampling locations are labelled 1 and 2 on the diagram of the Port (Figure 1). The first sampling location used was the security check area close to the beginning of the route through the Port of Dover. Here some of the drivers have their passports checked and all drivers slow down to find out if they are to be checked. Bluetooth can be detected at ranges up to 100 meters [3] without sophisticated equipment but it is heavily dependent on conditions. This area is a good location for Bluetooth monitoring because it is under cover (allowing some degree of signal reflection), drivers are driving slowly, the capture point is close to the vehicles and drivers usually have their windows open for passport checks. The only downside to capturing here is that there are 6 possible lanes for drivers to take, but most take the three central ones. During the capture period of four hours approximately 1200 vehicles passed through this area and 796 Bluetooth devices were registered. Some vehicles may have more than one Bluetooth device so the exact percentage of vehicles sampled is not trivial to ascertain but discarding all but one of multiple, time synchronised Bluetooth signals largely removes replication.

The second sampling location used was after the ticketing area when sixteen lanes funnel into three. This area captured much fewer Bluetooth signals than location 1 (125 vs 796) for a similar period, even though there are fewer lanes. The reasons for this included much fewer open windows, faster speeds, and a propensity to use the middle lane of three. When the unique Bluetooth addresses were compared we found a total of 104 examples of a Bluetooth device being found at both locations. Figure 16 shows the trip times registered for these 104 Bluetooth emitters that were captured at the start and finish of the trip through the Port of Dover. Visually it is apparent that there are general trends of increasing and decreasing trip time associated with queues and congestion interspersed with longer or shorter individual trips that could be explained by events such as security checks, vehicles parking to retrieve items from their boot or motor bikes/fast moving vehicles negotiating the circuit quicker.

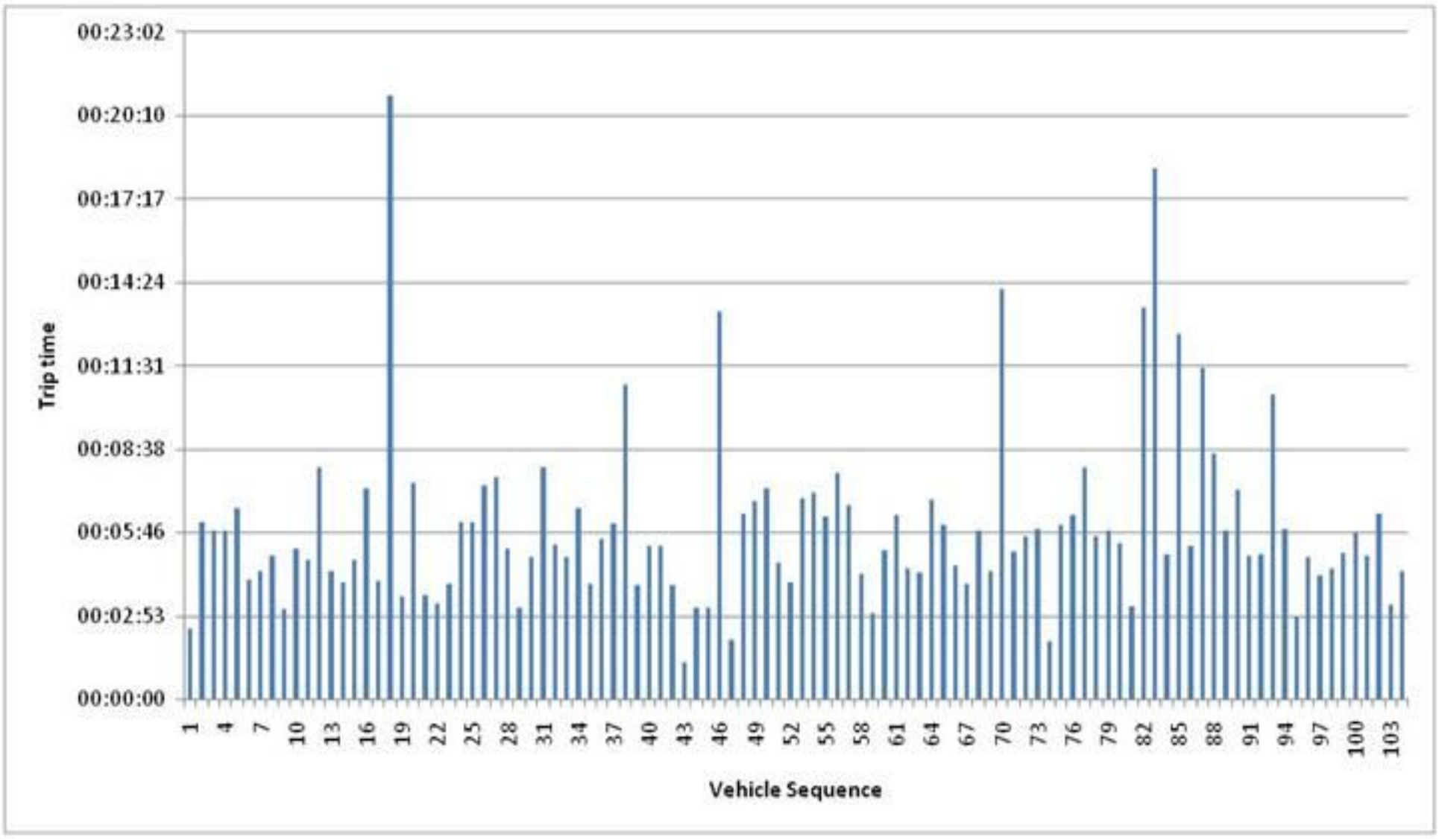

**Figure 16: Trip times captured by Bluetooth monitoring**

Arrival of anonymous vehicles at various locations was also captured using logging at the entry camera, the weighbridge and the ticketing kiosks. These captures were all partial, the video capture missed some vehicles because of obscuring, the weighbridge only captured RHVs and the ticketing was only available for a subset of the ferry operators. Using these real flows the current VISSIM simulation was stressed with a traffic flow that was an accurate representation of the real flow. Video cameras were also used to capture vehicle details at the same locations at the Bluetooth capture, enabling the derivation of trip times that can be compared with the Bluetooth derived trip times. Figure 17 shows how the Simulation compared with the Bluetooth capture and the visual inspection capture. Over 100 vehicles were simultaneously sampled for each example. The Bluetooth and visual capture are very closely correlated. The correlation between the simulation results and the capture results is less strong. Looking at the statistics for the three datasets (Table 2) it appears that the simulation produces slightly lower trip times but not to a statistically significant degree.

| | Trip time data (seconds) | | |
|---|---|---|---|
| | ***Simulation*** | ***Bluetooth*** | ***Camera*** |
| Mean | 319 | 358 | 343 |
| Median | 291 | 301 | 306 |
| Standard Deviation | 110 | 181 | 137 |
| Max | 729 | 1250 | 860 |

**Table 2: Means and Standard Deviations for Trip time measurements**

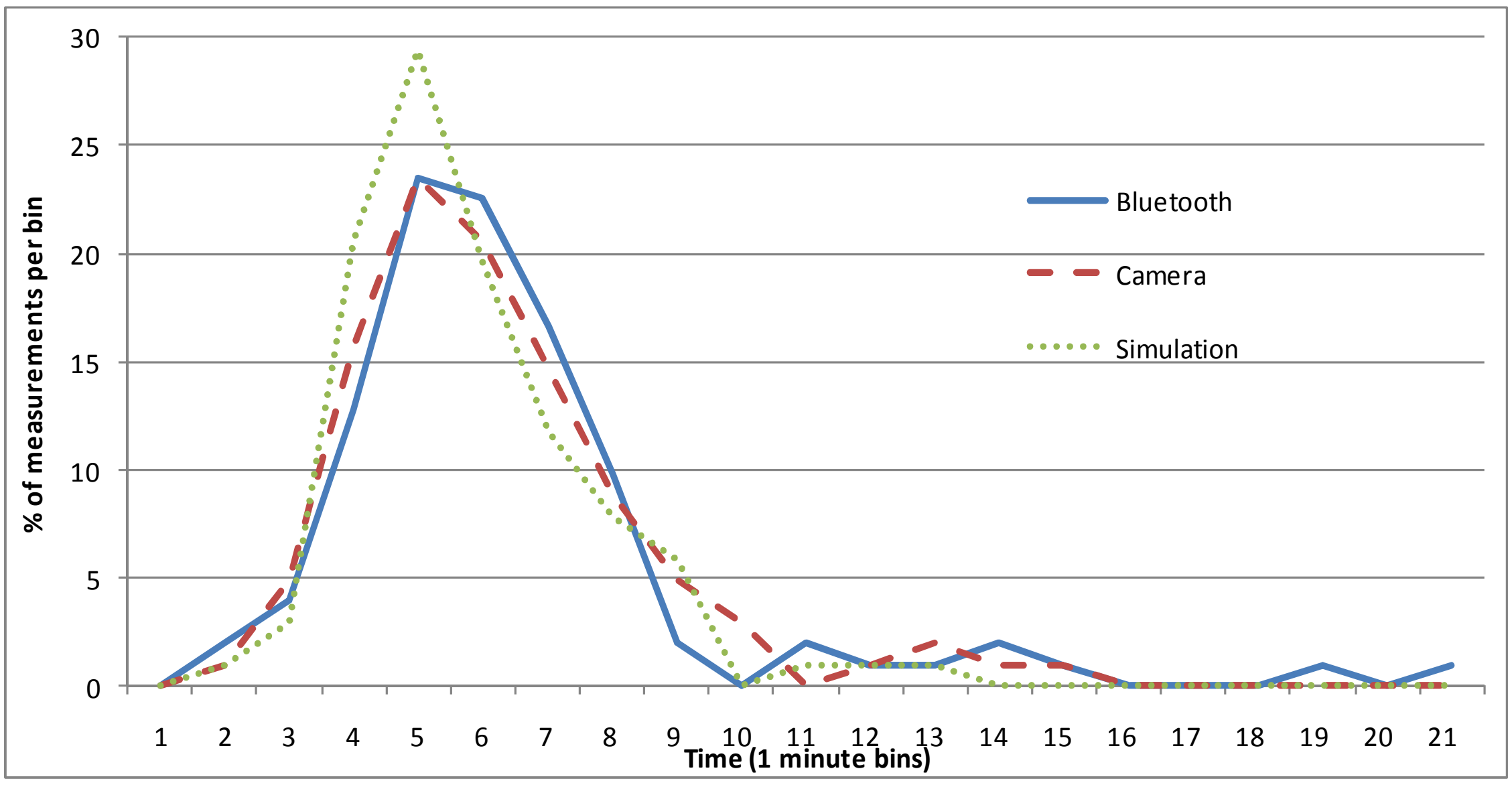

**Figure 17: Probability Density Function of Trip times as measured by Bluetooth and Visual Sampling**

The more complex a microsimulation is the more validation options there are. For the Dover simulation we could validate the trip times at the start and finish of the Dover route. It is important that the *simulation* shows similar trip times and trip statistics to *real* observations as shown in Figure 17 and Table 2. With complex microsimulations it is also important to validate individual components of the simulation. It is important to do this to ensure that correct gross statistics are not being achieved by aggregation of incorrect modelling of sub components. For instance, in some circumstances the summation of two incorrectly modelled sections may give accurate overall trip times but when the systems is stressed in different ways there are no assurances that the overall statistics will remain accurate.

## VI. CONCLUSIONS

A procedure for validation of microscopic traffic simulation models is tested, and its application to the simulation toolbox VISSIM is demonstrated. The validation efforts are performed at the microscopic level using Bluetooth monitoring and visual counting of vehicles. Analysis of variance of the simulation results versus the field data shows Bluetooth capture to be a useful approach for micro-monitoring of vehicles but care must be taken when choosing a collection site. While Bluetooth may not remain popular forever it is entirely plausible that some kind of detectible wireless protocol will continue to be available for the foreseeable future. The process detailed here may be considered a step towards the development of a systematic approach for validation of traffic simulation models. We have taken this work forward by using the refined model to assess the performance of modifications to the Port. We also plan to investigate how accurately a Microsimulation model can capture rare but important events, the key to this will be assuring that anomalous behaviour is not due to simulation construction.

When building a microsimulation great care must be taken to ensure each component is as accurate as possible as small errors in design can lead to disproportionately large errors. This is especially the case if actual behaviour is replaced with probabilistic approaches, while these can ensure representative statistics they can also introduce gross errors when coupled with strict lane discipline and can also be an example of overcalibration. Unlike areas such as Neural Networks and Statistics, overfitting or overcalibration of a simulation is much less well understood and therefore methods to avoid it are less well known.

There is a requirement in an agent based simulation to have appropriately intelligent agents that best reflect actual behaviour without introducing significant overheads in terms of complexity and hardware requirements. Having agents with representative behaviour reduces the need to overcalibrate the system by using popular methods such as probabilistic routing. This is evidenced when probabilistic routing is replaced by allowing drivers to make natural decisions on lane selection which results in much more consistent simulation performance at high flow levels.

Alternative methods for validation of simulation results are shown, with Bluetooth capture proving to be a viable, low maintenance method of trip time sampling. A more comprehensive Bluetooth capture may generate enough trip times to test the existing simulation at a more conclusive statistical level. Bluetooth is only one of many wireless protocols and may not be a long term standard for vehicle wireless comes but the approach should be equally effective with future wireless communication techniques.

## ACKNOWLEDGMENTS

We would like to thank the Port of Dover, particularly Daniel Gillett and Paul Simmons, for their valuable contributions, for supplying existing VISSIM models and for allowing access to locations within the Port. This research has been funded by the EPSRC.

## REFERENCES

[1] Charypar, D., K.W. Axhausen and K. Nagel (2007a) An event-driven parallel queue-based microsimulation for large scale traffic scenarios, paper presented at the 11th World Conference on Transportation Research, Berkeley, June 2007.

[2] Li, Ting, Eric van Heck, Peter Vervest, Jasper Voskuilen, Freek Hofker, and Fred Jansma. 2006. Passenger Travel Behavior Model in Railway Network Simulation. In Proceedings of the 2006 Winter Simulation Conference, eds. L. F. Perrone, F. P. Wieland, J. Liu, and B. G. Lawson, 1380-1387. CD ISBN 1-4244-0501-7, IEEE Catalog Number 06CH37826

[3] M. Fellendorf and P. Vortisch. "Validation of the microscopic traffic flow model VISSIM in different real-world situations". 80th Meeting of the Transportation Research Board. Washington, D.C., January, 2001.

[4] http://www.doverport.co.uk/?page=PortDevelopment

[5] Bluetooth SIG, Specification of the Bluetooth System, http://www.bluetooth.com, 2003.

[6] Do H. Nam, Donald R. Drew, Automatic measurement of traffic variables for intelligent transportation systems applications, Transportation Research Part B: Methodological, Volume 33, Issue 6, August 1999, Pages 437-457, ISSN 0191-2615,

[7] Davidsson, P., Henesey, L., Ramstedt, L., Tornquist, J. and Wernstedt, F. 2005. An analysis of agent based approaches to transport logistics. Transportation Research: Part C: Emerging Technologies, 13(4), 255–271.Hawkins, D. M. The problem of overfitting. J. Chem. Inf. Comput. Sci. 2004, 44, 1-12

[9] Smith, L., R. Beckman, D. Anson, K. Nagel, and M. E. Williams. 1995. Transims: Transportation analysis and simulationsystem. In Proceedings of the Fifth National Conference on Transportation Planning Methods. Seattle, Washington.

[10] Cameron, G., and G. Duncan. 1996. Paramics: Parallel microscopic simulation of road traffic. In Journal of SuperComputing

[11] Prevedouros, P. D., and Y. Wang. 1999. Simulation of a large freeway/arterial network with integration, tsis/corsim and watsim. In Transportation Research Record 1678, 197–207

[12] Concepts, I. T. 2001. Vissim simulation tool. In http://www.itc-world.com/VISSIMinfo.htm

[13] Sargent RG. Verifying and validating simulation models. 1996 Winter Simulation Conference, Coronado, CA,1996. p. 55–64.

[15] Tetko, I.V.; Livingstone, D.J.; Luik, A.I. Neural network studies. 1. Comparison of Overfitting and Overtraining, J. Chem. Inf. Comput. Sci., 1995, 35, 826-833

[16] D.E. Goldberg, K. Deb, A comparative analysis of selection schemes used in genetic algorithms, in: G.J.E. Rawlins (Ed.), Foundations of Genetic Algorithms, Morgan Kaufmann, San Mateo, CA, 1991, pp. 69–93.

[17] Stewart Robinson (2004). Simulation - The practice of model development and use. Wiley.